\documentclass{PoS}

\usepackage{graphicx}
\usepackage{amsmath}
\usepackage{amssymb}
\newcommand{\be}{\begin{equation}}
\newcommand{\ee}{\end{equation}}
\newcommand{\bea}{\begin{eqnarray}}
\newcommand{\eea}{\end{eqnarray}}
\newcommand{\bi}{\begin{itemize}}
\newcommand{\ei}{\end{itemize}}
\newcommand{\ben}{\begin{enumerate}}
\newcommand{\een}{\end{enumerate}}
\newcommand{\bt}{\begin{tabbing}}
\newcommand{\et}{\end{tabbing}}

\title{
   \begin{picture}(0,0)(0,0)%
   \end{picture}%
Phase structure of $N_{\rm f}=3$ QCD at finite temperature and density by Wilson-Clover fermions
}

\ShortTitle{
Phase structure of $N_{\rm f}=3$ QCD at finite temperature and density by Wilson-Clover fermions
}

\author{
   \speaker{Shinji Takeda}$^{a,b}$\thanks{E-mail: takeda@hep.s.kanazawa-u.ac.jp}, 
   Xiao-Yong Jin$^c$,
   Yoshinobu Kuramashi$^{b,d,e}$,
   Yoshifumi Nakamura$^b$,
   and
   Akira Ukawa$^b$
   \\
   \\
   \\
   \llap{$^a$}
Institute of Physics, Kanazawa University, Kanazawa 920-1192, Japan
   \\
   \llap{$^b$}
RIKEN Advanced Institute for Computational Science,
Kobe, Hyogo 650-0047, Japan
   \\
   \llap{$^c$}
Argonne Leadership Computing Facility, Argonne National Laboratory, Argonne, IL 60439, USA
   \\
   \llap{$^d$}
Graduate School of Pure and Applied Sciences,
University of Tsukuba,
Tsukuba, Ibaraki 305-8571, Japan
   \\
   \llap{$^e$}
Center for Computational Sciences,
University of Tsukuba,
Tsukuba, Ibaraki 305-8577, Japan
}
   
\abstract{
We investigate the phase structure of 3-flavor QCD in the
presence of finite quark chemical potential by using Wilson-Clover fermions.
To deal with the complex action with finite density, we adopt the phase reweighting method.
In order to survey a wide parameter region, we employ the multi-parameter reweighting method
as well as the multi-ensemble reweighting method.
Especially, we focus on locating the critical end point that characterizes the phase structure.  
It is estimated by the kurtosis intersection method for the quark condensate.
For Wilson-type fermions, the correspondence between bare parameters and
physical parameters is indirect,
thus we present
a strategy to transfer the bare parameter phase structure to the physical one.
We conclude that the curvature with respect to the chemical potential is positive.
This implies that, if one starts from a quark mass in the region of crossover at zero chemical potential,
one would encounter a first-order phase transition when one raises the chemical potential.
}

\FullConference{The 33rd International Symposium on Lattice Field Theory\\
		 14 -18 July  2015\\
		 Kobe International Conference Center, Kobe, Japan}

\begin{document}

\section{Introduction}

The location of the critical end point in the QCD phase diagram at finite density is an important 
unsolved issue.
In this article, we address the  issue of how the critical end line extends when switching on the quark chemical potential.
An interesting result was reported in \cite{deForcrand:2006pv,deForcrand:2007rq} which explored the imaginary chemical potential approach with the naive staggered fermion action. 
There it was reported that the critical surface has a negative curvature in the $\mu$ direction. This means that a first-order phase transition at zero chemical potential disappears when the chemical potential is increased, rather contrary to one's naive guess.  
Our purpose in this report is to study this question by simulations with real chemical potential using the Wilson-clover fermion action.  
The detail of this article can be seen in \cite{Jin:2015taa}.

\section{Strategy}
\label{sec:strategy}
In this section, we explain our strategy to survey the phase space for $N_{\rm f}=3$ QCD, especially how to identify the critical end point for the Wilson-type fermions and how to obtain the curvature of the critical end line on the $\mu$-$m_{\pi}$ plane.
Note that in this section we do not use lattice units when expressing dimensionful physical quantities.

\begin{figure}[t]
\begin{center}
\begin{tabular}{cc}
\hspace{-12mm}
%\vspace{8mm}
\scalebox{0.35}{\includegraphics{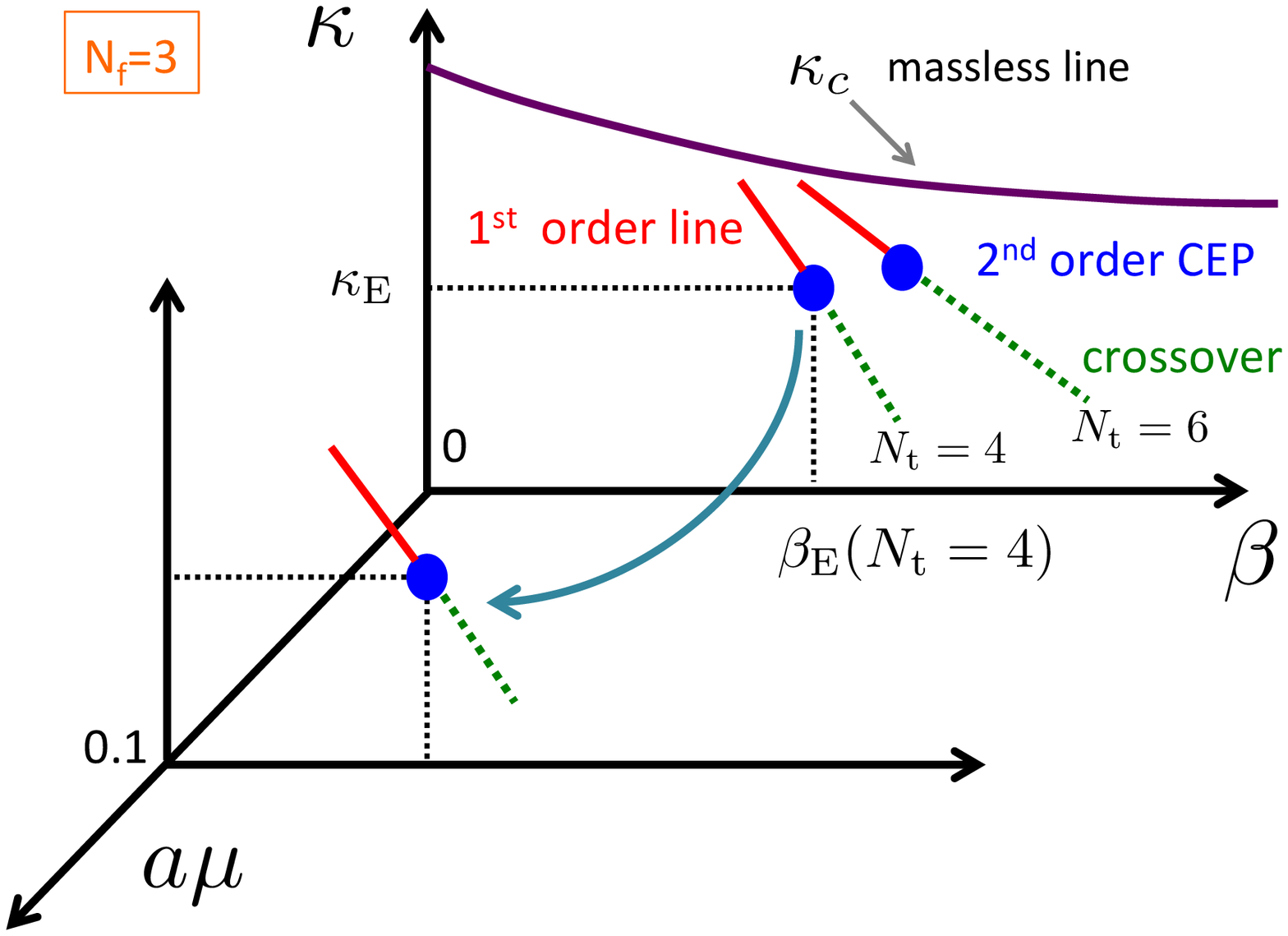}}
&
\hspace{-13mm}
\scalebox{0.35}{\includegraphics{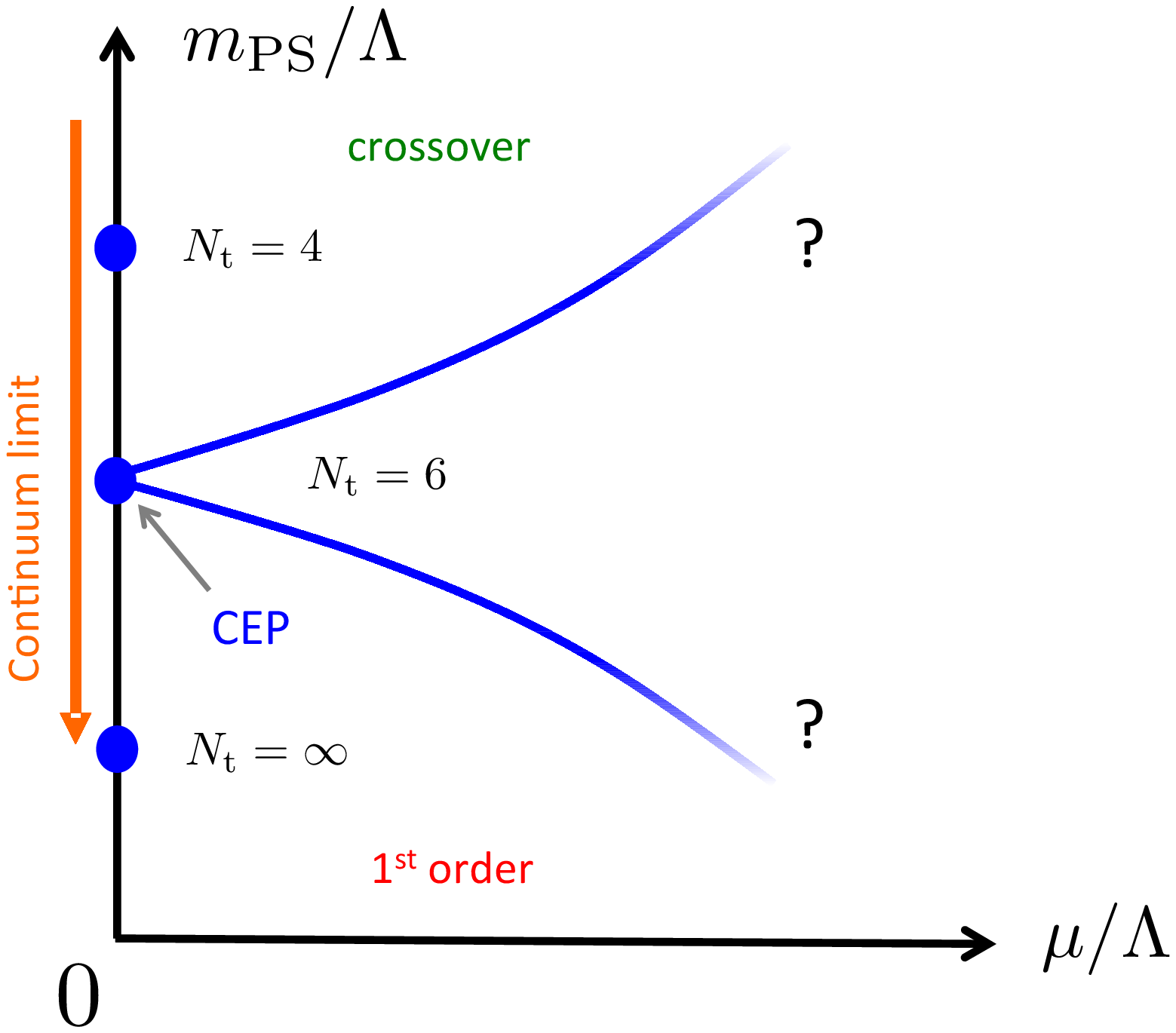}}
\\
\end{tabular}
\end{center}
\vspace{-8mm}
\caption{
Strategy:
The left panel is the phase diagram for bare parameters spanned by $\beta$, $\kappa$ and $a\mu$ for $N_{\rm f}=3$.
The right panel is the same phase diagram but depicted for physical parameters spanned by $m_{\rm PS}/\Lambda$ and $\mu/\Lambda$ where $\Lambda$ is some reference physical quantity at zero density.
The blue line is the critical line.
We study the signature of the curvature of critical line with fixed $N_{\rm t}=6$.
}
\label{fig:strategy}
\end{figure}

First we consider the zero density case.
Since the quark masses are all degenerate, we have only two bare parameters $\beta$ and $\kappa$ ($a\mu=0$ plane in the left panel in Fig.~\ref{fig:strategy}).
For a given temporal lattice size, say $N_{\rm t}=4$, by using the peak position of susceptibility of quark condensate, one can draw the line of  finite temperature transition.
The transition changes from being of first order to cross over at a second order critical end point. 
We compute the kurtosis of quark condensate along the transition line for a set of spatial volumes.
The intersection point is identified as the critical end point \cite{Karsch:2001nf}.
In this way, we can determine the critical end point in the bare parameter space
$(\beta_{\rm E},\kappa_{\rm E})$ and this procedure can be repeated for other values of $N_{\rm t}$.

%%%%%%%%
%%%%%%%%%

In order to translate the critical end point in the bare parameter space to that in the physical parameter space, we measure dimensionless ratios of pseudo-scalar meson mass and some reference quantity with mass-dimension one $m_{\rm PS}/\Lambda$ for the bare parameters $(\beta_{\rm E},\kappa_{\rm E})$ by a zero temperature simulation.
%One can choose any reference quantity $\Lambda$.
To avoid the multiplicative renormalization, we  use $m_{\rm PS}$ in the numerator of the ratio and not quark masses.
In this way we pin down the critical end point in the physical parameter space. % spanned by $m_{\rm PS}/\Lambda$ and $\mu/\Lambda$. 
By repeating the same calculation for increasingly larger values of $N_{\rm t}$, we can take the continuum limit of the critical end point (this strategy is in fact used in our zero density study \cite{Jin:2014hea}),
\begin{equation}
\frac{m^{\rm cont}_{\rm PS,E}(\mu=0)}{\Lambda^{\rm cont}_{\rm E}(\mu=0)}
=
\lim_{N_{\rm t}\rightarrow\infty}
\frac{m_{\rm PS,E}(\mu=0)}{\Lambda_{\rm E}(\mu=0)}.
\end{equation}

When switching on the chemical potential, the basic procedure is the same;  one just has to repeat the same analysis on a different plane with $\mu\neq0$ (see the left panel in Fig.~\ref{fig:strategy}).
For a fixed lattice temporal size, $N_{\rm t}=6$,
in order to draw the critical end line, we consider a pair of dimensionless ratios
\begin{equation}
\frac{m_{\rm PS,E}(\mu)}{m_{\rm PS,E}(0)}
\hspace{10mm}
\mbox{and}
\hspace{10mm}
\frac{\mu}{T_{\rm E}(0)},
\label{eqn:ratios}
\end{equation}
where for each ratio we have chosen proper reference quantities at zero density.
By plotting these two quantities one can obtain a critical line as shown in the right panel of Fig.~\ref{fig:strategy}.
Especially, we are interested the curvature $\alpha_1$ of the fitting form
\begin{equation}
\left(
\frac{m_{\rm PS,E}(\mu)}{m_{\rm PS,E}(0)}
\right)^2
=
1
+
\alpha_1
\left(
\frac{\mu}{\pi T_{\rm E}(0)}
\right)^2
+
\alpha_2
\left(
\frac{\mu}{\pi T_{\rm E}(0)}
\right)^4
+....
\label{eqn:MPSmuT}
\end{equation}

\section{Simulation details}
\label{sec:details}

We employ the Wilson-clover fermion action with non-perturbatively tuned $c_{\rm sw}$~\cite{CPPACS2006} in the presence of chemical potential. % with the anti-periodic boundary condition in the temporal direction for fermion fields while the periodic boundary condition is imposed for spatial direction.
The Iwasaki gauge action \cite{iwasaki} is used for the gluon sector. % and gauge link variables satisfy the periodic boundary condition.
The number of flavor is three, $N_{\rm f}=3$, and the masses and chemical potentials for quarks are all degenerate.
The temporal lattice size and the simulated quark chemical potential are fixed to $N_{\rm t}=6$ and $a\mu=0.1$, respectively. %, and thus $\mu/T=0.6$. 
In our study, the phase reweighting method is used to deal with the complex phase.  
To survey a wide range of $\mu$ and $\kappa$, we adopt the multi-parameter reweighting method.
To perform  finite size scaling analysis, the spatial volume is changed over the linear sizes $N_{\rm s}=8$, $10$ and $12$.
In order to search for the transition point, we select four $\beta$ points ($\beta=1.70$, $1.73$, $1.75$ and $1.77$) and for each $\beta$, we vary $\kappa$ to locate the transition point.

Configurations are generated by RHMC %\cite{RHMC} 
with the phase quenched quark determinant.
%The MD step size is chosen such that a reasonable acceptance rate $\gtrsim80\%$ is retained.
For each lattice parameter set $(\beta,\kappa,N_{\rm t},N_{\rm s})$ we generate $O(100,000)$ trajectories, with the configurations stored at every 10th trajectory. %;  the order of number of configurations are $O(10,000)$ for each parameter set.
The phase factor is computed exactly using the analytical reduction technique \cite{Danzer:2008xs,Takeda:2011vd,takedanote} for all stored configurations.
The dense matrix obtained by the reduction is numerically computed on GPGPU with LAPACK routines.
We measure the trace of quark propagator and its higher power up to fourth order which are used not only for the computation of higher moments of quark condensate but also for the parameter reweighting.
In the computation of traces, we adopt the noise method with 20 Gaussian noises that is checked to be sufficient to control the noise error.

For each fixed parameter set ($\beta$, $a\mu$, $N_{\rm t}$, $N_{\rm s}$), we make runs at several values of $\kappa$.
In order to integrate those runs we adopt the multi-ensemble reweighting technique~\cite{reweighting} and search for the transition point in $\kappa$ for the fixed parameter set.
%See Appendix \ref{sec:multiensemblereweighting} for the details of the multi-ensemble reweighting.
We employ some approximation, checking its validity numerically,  to efficiently evaluate the quark determinant in the reweighting factor as well as observables at many reweighting points.

%In our approach, there are practically two important issues: the overlap problem and the validity of approximation
%made at calculating the ratio of quark determinant in the reweighting factor.
%The issue of the overlap problem will be addressed in the next section.
%The validity of the approximation is discussed in Appendix \ref{sec:reweighting} and the conclusion is that the approximation we made is safe in our parameter region.

\section{Simulation results}
\label{sec:results}

%%%%%%%%
\begin{figure}[t]
\begin{center}
\begin{tabular}{cc}
%\hspace{-10mm}
\scalebox{0.8}{\includegraphics{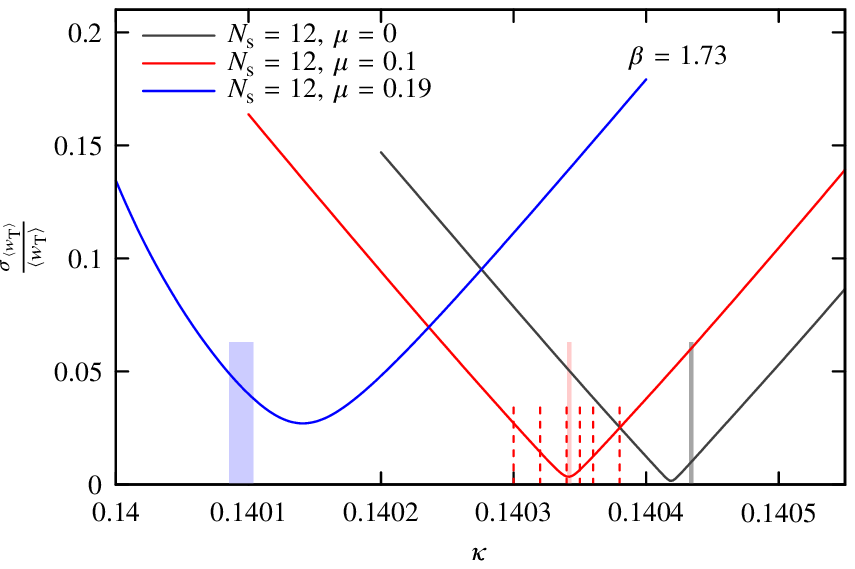}}
&
%\hspace{-10mm}
\scalebox{0.65}{\includegraphics{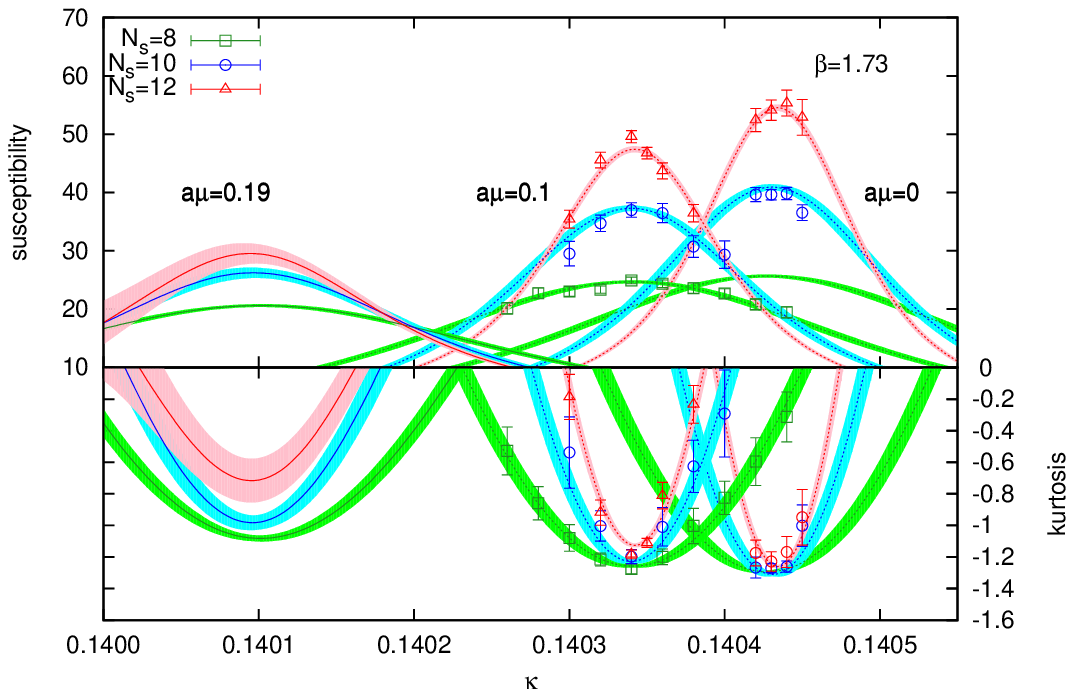}}
\end{tabular}
\end{center}
\caption{
The left panel is the relative error of the reweighting factor of multi-ensemble reweighting as a function of $\kappa$ for $1.73$.
Here only three selected values of the chemical potential are shown for each $\beta$.
The spatial lattice is fixed $N_{\rm s}=12$.
%Note that the y-axis is logarithmic scale.
The right panel is the susceptibility and kurtosis of quark condensate as a function of $\kappa$ for $1.73$.
In $\beta=1.73$ with $N_{\rm s}=10,12$, we also plot the raw data at $a\mu=0$ given in the zero density study \cite{Jin:2014hea}.
%These raw data and the reweighting results are consistent with each other, although they are completely independent.
%This shows that the reweighting together with the approximation used for the reweighting factors and observables is fine.
}
\label{fig:reweighting}
\end{figure}
%%%%%%%%%

The relative error of the reweighting factor of the multi-ensemble reweighting is plotted in Fig. \ref{fig:reweighting} (left panel).
The relative error takes small values even at a large chemical potential $a\mu\approx0.2$, and 
the reweighting factor is significantly away from zero beyond several sigmas.
Thus we conclude that the overlap problem is not so severe in our parameter region.

%%%%%%%%
\begin{figure}[t]
\begin{center}
\begin{tabular}{c}
\scalebox{0.5}{\includegraphics{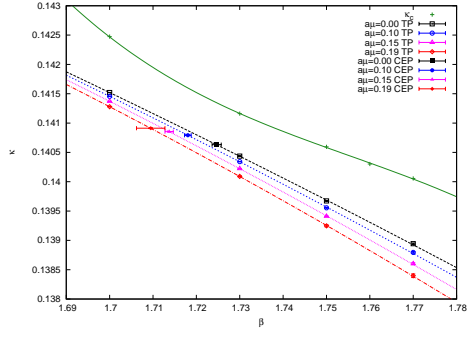}}
\end{tabular}
\end{center}
\caption{
The phase diagram of $N_{\rm f}=3$ QCD with finite chemical potential projected on the $(\beta,\kappa)$ plane.
The transition points are expressed by open symbol while the critical end points are given by filled one.
The lower $\beta$ side of the critical end point is the first order phase transition region. 
%For larger chemical potential, the critical point moves toward the upper-left corner.
The $\kappa_{\rm c}$ line where the pion mass vanishes is also shown.
}
\label{fig:betakappa}
\end{figure}
%%%%%%%%%

The right panel of Fig.~\ref{fig:reweighting} shows curves of the susceptibility and kurtosis for quark condensate obtained by the multi-ensemble reweighting. 
For $a\mu=0.1$, the averages at each point of data generation are shown in order to illustrate how multi-ensemble curves interpolate those raw data. 
At $\beta=1.73$, the curves reweighted to $a\mu=0$ can be compared with data generated at zero density \cite{Jin:2014hea}.  The agreement supports the validity of multi-ensemble reweighting and jackknife error estimation away from $a\mu=0.1$. 
The applicable range of $\mu/\kappa$-reweighting depends on $\beta$, and judged from the growth of error, the lower $\beta$ tends to have a larger applicable range. 
As seen in the figure, the locations of the maximum of susceptibility and minimum of kurtosis are consistent with each other.
%Furthermore the skewness zero location is also consistent with them although it is not shown here.
We take the location of the maximum of susceptibility as the transition point.
%The numerical values are summarized in Table~\ref{tab:transitionpoint} where the peak hight of susceptibility $\chi_{\rm max}$ and the minimum of kurtosis $K_{\rm min}$ are also listed for selected values of $a\mu$.
We observe that the volume dependence of the transition points is rather mild. Hence the thermodynamic limit can be safely taken with a fitting ansatz,
$
\kappa_{\rm t}(N_{\rm s})=\kappa_{\rm t}(\infty)+c/N_{\rm s}^3.
$
The phase diagram of bare parameters $\beta$ and $\kappa$ is given in Fig.~\ref{fig:betakappa}.
%The transition lines have a sensitivity on the value of chemical potential, $a\mu$.

%%%%%%%%%
\begin{figure}[t]
\begin{center}
\begin{tabular}{ccc}
%\hspace{-5mm}
\scalebox{.6}{\includegraphics{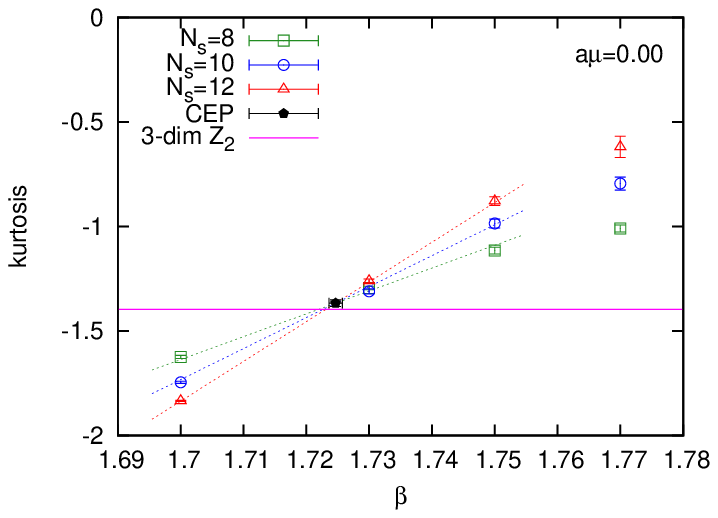}}
&
\scalebox{.6}{\includegraphics{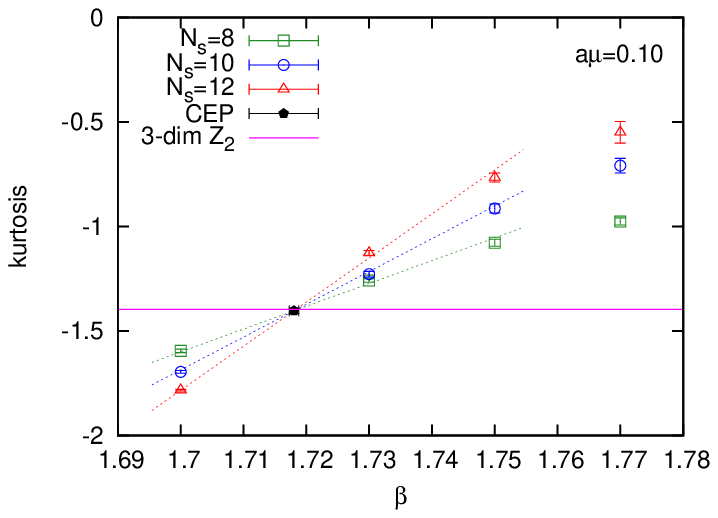}}
&
\scalebox{.6}{\includegraphics{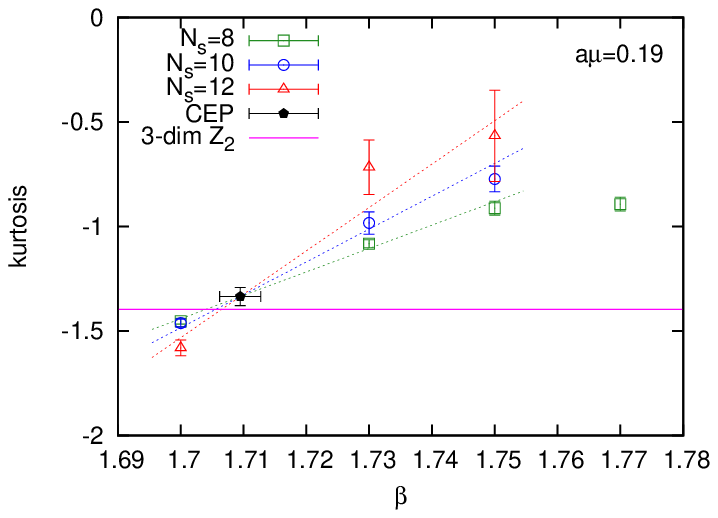}}
\\
\end{tabular}
\end{center}
\caption{
Kurtosis intersection at $a\mu=0.00-0.19$.
In the fitting, three lowest values of $\beta$ are used.
The black pentagon represents the critical end point (CEP) in bare parameter space and it moves to lower side for larger chemical potential.
%See Table \ref{tab:kurtosisintersectionfit} for the values of fitting parameters.
The horizontal line shows $K_{\rm E}=-1.396$ for 3-D ${\rm Z}_2$ universality class.
%In this region of the chemical potential, the value of $K_{\rm E}$ is constant, namely the universality class does not change.
}
\label{fig:krtintersection}
\end{figure}

The next step is to determine the critical end point.
For that purpose we adopt the kurtosis intersection method \cite{Karsch:2001nf}.
Figure~\ref{fig:krtintersection} plots the minimum of kurtosis as a function of $\beta$ for some selected values of $a\mu$.
This shows that a strong first order phase transition at lower $\beta$ becomes weaker for higher $\beta$ and such a change becomes rapid for larger volumes.
We fit the data with the fitting form \cite{deForcrand:2006pv} inspired by finite size scaling,
$
K_{\rm min}
=
K_{\rm E}
+
A N_{\rm s}^{1/\nu}(\beta-\beta_{\rm E}),
$
where $K_{\rm E}$, $A$, $\nu$ and $\beta_{\rm E}$ are fitting parameters.
We observe that the resulting exponent $\nu$ and the value of kurtosis at the critical end point $K_{\rm E}$ are independent of $a\mu$ within errors,  and they are consistent with the values of 3-D ${\rm Z}_2$ universality class, $\nu=0.63$ and $K_{\rm E}=-1.396$ respectively.
On the other hand, the universality class of 3-dimensional O(2) and 3-dimensional O(4) are rejected, rather strongly by the value of $K_E$. 
We superimpose the obtained critical end points $(\beta_{\rm E}(a\mu),\kappa_{\rm E}(a\mu))$ for $0\le a\mu \le0.19$ in the phase diagram of Fig.~\ref{fig:betakappa}.
%The critical end point moves toward the upper-left corner by increasing $a\mu$.

\section{Critical line and its curvature}
\label{sec:criticalline}

The analysis of the critical line below requires a careful manipulation with scale setting.  Thus we distinguish quantities in lattice units from those  in physical units by placing tilde on the former, {\it e.g.,} chemical potential in physical units is denoted as $\mu$ and that in lattice units by  $\tilde\mu=a\mu$. 

In the previous section, we have determined the critical end points in the bare parameter space.
We now translate the critical end point on the $(\beta,\kappa)$ plane to the physical space, to obtain the critical line as one varies $\mu$, and  finally to extract its curvature.
For that purpose, as explained in Sect.~\ref{sec:strategy}, we need to compute the pair of ratios in eq.(\ref{eqn:ratios}) as follows,
\be
\frac{m_{\rm PS,E}(\mu)}{m_{\rm PS,E}(0)}
=
\frac{\tilde m_{\rm PS,E}(\tilde\mu)}{\tilde m_{\rm PS,E}(0)}
\cdot
\frac{a(0)}{a(\tilde\mu)},
%\label{eqn:A},
\hspace{10mm}
\frac{\mu}{T_{\rm E}(0)}
=
\tilde\mu
\cdot
\frac{a(0)}{a(\tilde\mu)}
\cdot
N_{\rm t},
\label{eqn:B}
\ee
where $\tilde m_{\rm PS,E}(\tilde\mu)$ is the pseudo-scalar (PS) meson mass in lattice units evaluated at $(\beta,\kappa)=(\beta_{\rm E}(\tilde\mu),\kappa_{\rm E}(\tilde\mu))$.
%Note that the PS mass at the critical point does not depend on $\tilde\mu$ directly, but only through $\beta_{\rm E}$ and $\kappa_{\rm E}$ at $\tilde\mu$.
The PS mass is measured by the zero temperature simulation at $\beta_{\rm E}$ and $\kappa_{\rm E}$.
On the other hand, the lattice spacing requires some careful thought as follows.

We usually determine the lattice spacing by choosing a line of constant physics (LCP) and specifying the value of a dimensionful physical quantity on that line.  For example, one may choose the dimensionless combination $m_{\rm PS}\sqrt{t_0}$ for specifying the LCP, and the value of
%$\sqrt{t_0}$
$m_{\rm PS}$
in physical units to determine the lattice spacing along the chosen LCP,
\be
a(\beta,y)
=
%\frac{1/\sqrt{\tilde t_0}(\beta,\kappa_y(\beta))}{1/\sqrt{t_0}(y)}.
\frac{\tilde m_{\rm PS}(\beta,\kappa_y(\beta))}{m_{\rm PS}(y)},
\label{eqn:latticespacingLCP}
\ee
where $y$ is the value of the constant physics $y=m_{\rm PS}\sqrt{t_0}$ and $\kappa_y(\beta)$ is defined such that
$
y=\tilde m_{\rm PS}(\beta,\kappa_y(\beta)) \sqrt{\tilde t_0}(\beta,\kappa_y(\beta))
$
holds for each $\beta$.
The notation of the lattice spacing in eq.(\ref{eqn:B}) means that
$
a(\tilde\mu)
=
a(\beta_{\rm E}(\tilde\mu),y).
$
%Note that, again, the lattice spacing does not depend on $\mu$ directly, but only though the $\beta_{\rm E}$ at $\tilde\mu$.
Along LCP, where the physical unit mass in the denominator in eq.(\ref{eqn:latticespacingLCP}) is not known a priori but common,
the physical mass cancels out in the ratio of lattice spacings %$a(0)/a(\tilde\mu)$
and the ratio may be computed by using the PS mass in lattice units,
\be
\frac{a(0)}{a(\tilde\mu)}
=
\frac
%{1/\sqrt{\tilde t_0}(\beta_{\rm E}(0),\kappa_{\rm ref}(\beta_{\rm E}(0)))}
%{1/\sqrt{\tilde t_0}(\beta_{\rm E}(\tilde\mu),\kappa_{\rm ref}(\beta_{\rm E}(\tilde\mu)))},
{\tilde m_{\rm PS}(\beta_{\rm E}(0),\kappa_y(\beta_{\rm E}(0)))}
{\tilde m_{\rm PS}(\beta_{\rm E}(\tilde\mu),\kappa_y(\beta_{\rm E}(\tilde\mu)))}.
\ee
%where $	1/\sqrt{\tilde t_0}$ is the Wilson flow scale in lattice units.
In the following, for the computation of the ratio of lattice spacings,
we use the Wilson flow scale instead of the PS mass since the former is more precisely calculated
\be
\frac{a(0)}{a(\tilde\mu)}
=
\frac
%{1/\sqrt{\tilde t_0}(\beta_{\rm E}(0),\kappa_{\rm ref}(\beta_{\rm E}(0)))}
%{1/\sqrt{\tilde t_0}(\beta_{\rm E}(\tilde\mu),\kappa_{\rm ref}(\beta_{\rm E}(\tilde\mu)))},
{1/\sqrt{\tilde t_0}(\beta_{\rm E}(0),\kappa_y(\beta_{\rm E}(0)))}
{1/\sqrt{\tilde t_0}(\beta_{\rm E}(\tilde\mu),\kappa_y(\beta_{\rm E}(\tilde\mu)))}.
\ee
%where $	1/\sqrt{\tilde t_0}$ is the Wilson flow scale in lattice units.

One can employ a different LCP by specifying a different value of $y^\prime(\neq y)$.
%$$m_{\rm PS}\sqrt{t_0}$.
The resulting lattice spacing coincides with that from the original ($y$) definition if, in specifying the value of the dimensionful quantity, one takes into account  the variation of that quantity in moving from the original LCP to a new LCP.  In general the agreement will not be exact due to scaling violations.
Thus differences one may observe in physical results due to the choice of LCP is a scaling violation effect. 
In the following, we choose two values for the line of constant physics, namely,
$
y=m_{\rm PS}\sqrt{t_0}=0.55 \mbox{  and  } 0.65.
%m_{\rm PS}/m_{\rm V}=0.67, 0.74.
$

\begin{figure}[t]
\begin{center}
\scalebox{.7}{\includegraphics{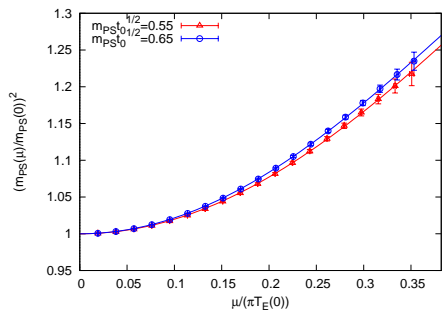}}
%\scalebox{1.2}{\includegraphics{t0_MPSMV_CL_JK_presentation.eps}}
\caption{
Critical line for constant physics
$y=m_{\rm PS}t_0^{1/2}=0.55, 0.65$
%$m_{\rm PS}/m_{\rm V}=0.67, 0.74$
with scale input: the Wilson flow scale.
%The both lines are bending toward upper direction.
\label{fig:criticalline}
}
\end{center}
\end{figure}

We use the Wilson flow scale and the hadron mass computed in Appendix A of Ref.~\cite{Jin:2014hea}.
By combining the above scale inputs and the information of critical end point at finite chemical potential determined in the previous section, we calculate the two ratios in eq.(\ref{eqn:B}).  
The results are plotted in Fig.~\ref{fig:criticalline}.
We extract the curvature by using the fitting form in eq.(\ref{eqn:MPSmuT}). 
We find the curvature values as given by
\be
\alpha_1=
\left\{
\begin{array}{cc}
1.924(60) & \mbox{ for } y=0.55,\\
2.148(39) & \mbox{ for } y=0.65.\\
\end{array}
\right.
\ee
We observe that the critical line has a sensitivity on the value of $y$.
This difference is considered as a systematic uncertainty caused by the choice of the scale setting as discussed above.
All in all, we find the curvature of the critical line to be positive with a statistical error of about 3\% and a systematic error of about 10\%.

\if0

%%%%%%%%%
\begin{table}[t]
\begin{center}
\begin{tabular}{l|l|l|l}
\hline\hline
constant physics
&
scale input
&
$\alpha_1$
&
$\alpha_2$
\\
\hline
$m_{\rm PS}\sqrt{t_0}=0.55$
&
$1/\sqrt{t_0}$
&
$1.924 ( 60 )$
&
$-0.58 ( 72 ) $
\\
$m_{\rm PS}\sqrt{t_0}=0.65$
&
$1/\sqrt{t_0}$
&
$2.148 ( 39 )$
&
$-1.74 ( 52 )$
\\
%\hline
%$m_{\rm PS}/m_{\rm V}=0.67$
%&
%$1/\sqrt{t_0}$
%&
%$1.758 ( 74 )$
%&
%\phantom{-}
%$0.21 ( 89 )$
%\\
%$m_{\rm PS}/m_{\rm V}=0.74$
%&
%$1/\sqrt{t_0}$
%&
%$2.032 ( 49 )$
%&
%$-1.15 ( 62 )$
%\\
\hline\hline
\end{tabular}
\end{center}
\caption{
The curvature of critical line in the fitting form in eq.(\ref{eqn:MPSmuT}) for scale inputs ($\sqrt{t_0}$) and the values of constant physics
$m_{\rm PS}\sqrt{t_0}=0.55,0.65$.
%$m_{\rm PS}/m_{\rm V}=0.67,0.74$.
The error of curvature is estimated by jackknife method.
\label{tab:curvature}
}
\end{table}
%%%%%%%%%

\fi

%// Acknowledgments -----------------------------------------------------------

\vspace{3mm}
This research used computational resources of the K computer provided by the RIKEN Advanced Institute for Computational Science
through the HPCI System Research project (Project ID:hp120115), and the HA-PACS
provided by Interdisciplinary Computational Science Program
in Center for Computational Sciences, University of Tsukuba.
This work is supported by JSPS KAKENHI %Grants-in-Aid for Young Scientists B 
Grant Numbers 23740177 and 26800130.
This work was supported by FOCUS Establishing Supercomputing Center of Excellence.

\end{document}